\documentclass[preprintnumbers,article,amsmath,amssymb,floatfix,10pt,prd,twocolumn,
superscriptaddress,nofootinbib]{revtex4-2}
\usepackage{bm}
\usepackage{amsfonts}
\usepackage{graphicx}
\usepackage{latexsym}
\usepackage[latin1]{inputenc}
\usepackage{graphicx}
\usepackage{amsmath}
\usepackage{palatino}
\usepackage{mathpazo}
\usepackage{textcomp}
\usepackage{float}
\usepackage{booktabs}
\usepackage{dcolumn}
\usepackage{ragged2e}
\usepackage{hyperref}
\hypersetup{colorlinks,citecolor=blue}
\hypersetup{colorlinks=true,linkcolor=red,filecolor=magenta,    urlcolor=blue}
\usepackage{amsmath}
\usepackage{xcolor}
\usepackage{orcidlink}
\usepackage{epsfig}
\usepackage{caption}
\usepackage{subcaption}
\usepackage{commath}
\def\jnl@style{\it}
\def\aaref@jnl#1{{\jnl@style#1}}

\def\aaref@jnl#1{{\jnl@style#1}}

\def\aj{\aaref@jnl{AJ}}                   
\def\apj{\aaref@jnl{ApJ}}                 
\def\apjl{\aaref@jnl{ApJ}}                
\def\apjs{\aaref@jnl{ApJS}}               
\def\apss{\aaref@jnl{Ap\&SS}}             
\def\aap{\aaref@jnl{A\&A}}                
\def\aapr{\aaref@jnl{A\&A~Rev.}}          
\def\aaps{\aaref@jnl{A\&AS}}              
\def\mnras{\aaref@jnl{Mon.~Not.~Roy.~Astron.~Soc.}}             
\def\prd{\aaref@jnl{Phys.~Rev.~D}}        
\def\prc{\aaref@jnl{Phys.~Rev.~C}}  
\def\prl{\aaref@jnl{Phys.~Rev.~Lett.}}    
\def\qjras{\aaref@jnl{QJRAS}}             
\def\skytel{\aaref@jnl{S\&T}}             
\def\ssr{\aaref@jnl{Space~Sci.~Rev.}}     
\def\zap{\aaref@jnl{ZAp}}                 
\def\nat{\aaref@jnl{Nature}}              
\def\aplett{\aaref@jnl{Astrophys.~Lett.}} 
\def\apspr{\aaref@jnl{Astrophys.~Space~Phys.~Res.}} 
\def\physrep{\aaref@jnl{Phys.~Rep.}}      
\def\physscr{\aaref@jnl{Phys.~Scr}}       
\def\commat{\aaref@jnl{Comm.~Math.~Phys.}}              
\def\science{\aaref@jnl{Science}}               
\def\cqg{\aaref@jnl{Classical Quant.~Grav.}}            
\def\jpcs{\aaref@jnl{JPCS}}                                     
\def\ijmpd{\aaref@jnl{Int.~J.~Mod.~Phys.~D}}                    
\def\grg{\aaref@jnl{Gen.~Relat.~Gravit.}}               
\def\rpp{\aaref@jnl{Rep.~Prog.~Phys.}}          
\def\npa{\aaref@jnl{Nucl.~Phys.~A}}        
\def\lrr{\aaref@jnl{Living Rev.~Rel.}}                   
\def\jcap{\aaref@jnl{J.~Cosmology Astropart.~Phys.}}    
\def\rmp{\aaref@jnl{Rev.~Mod.~Phys.}}   
\def\epjc{\aaref@jnl{Eur.~Phys.~J.~C}} 
\def\plb{\aaref@jnl{~Phy.~Lett.~B}} 
\def\mpla{\aaref@jnl{Mod.~Phy.~Lett.~A}} 
\def\arxiv{\aaref@jnl{arxiv.org}}

\allowdisplaybreaks[1]

\begin{document}
\color{black}       
%

\title{Reconstruction of the scalar field potential in nonmetricity gravity through Gaussian processes}

\author{Gaurav N. Gadbail\orcidlink{0000-0003-0684-9702}}
\email{gauravgadbail6@gmail.com}
\affiliation{Department of Mathematics, Birla Institute of Technology and Science-Pilani, Hyderabad Campus, Hyderabad-500078, India.}

\author{Sanjay Mandal\orcidlink{0000-0003-2570-2335}}
\email{sanjaymandal960@gmail.com}
\affiliation{Faculty of Symbiotic Systems Science, Fukushima University, Fukushima 960-1296, Japan.}

\author{P.K. Sahoo\orcidlink{0000-0003-2130-8832}}
\email{pksahoo@hyderabad.bits-pilani.ac.in}
\affiliation{Department of Mathematics, Birla Institute of Technology and Science-Pilani, Hyderabad Campus, Hyderabad-500078, India.}

\author{Kazuharu Bamba \orcidlink{0000-0001-9720-8817}}
\email{bamba@sss.fukushima-u.ac.jp}
\affiliation{Faculty of Symbiotic Systems Science, Fukushima University, Fukushima 960-1296, Japan.}

%

\begin{abstract}
\textbf{Abstract :-} The accelerated expansion of the universe has been widely confirmed, posing challenges to the standard $\Lambda$CDM model, particularly the cosmological coincidence problem. This has motivated the exploration of modified gravity theories, including nonmetricity gravity, which explains cosmic acceleration without dark energy. In this work, we incorporate a quintessence scalar field into the nonmetricity framework to model both inflation and late-time acceleration. Employing the Gaussian process method with a square exponential kernel, we reconstruct the scalar field potential, $V(\phi)$, from observational Hubble data sets coming from cosmic chronometers (CC) as well as from the method of radial baryon acoustic oscillations (BAO) in a model-independent approach. This approach allows us to obtain a suitable quintessence scalar field model that aligns with the observational Hubble data under the framework of power-law nonmetricity gravity. Additionally, we compare our reconstructed potential with power-law scalar field potentials, revealing that these models show better agreement with the observational data, providing new insights into the dynamics of the universe. In contrast, we find that the early dark energy has minimal effect on the present-time accelerated expansion of the universe.


\end{abstract}

\maketitle

\date{\today}

\section{\NoCaseChange{Introduction}}
The discovery of the accelerated expansion of the universe has been confirmed by numerous independent observations \citep{A1,A2,A3,A4,A5,A6,A7,A8,A9,A10}, and it remains a subject of active investigation within the framework of various gravitational theories. Dark energy, proposed within the context of general relativity (GR) to explain this accelerated expansion \citep{DE1,DE2}, has taken on multiple forms. The most straightforward and widely used model for dark energy is the cosmological constant, which corresponds to vacuum energy at the cosmological scale. This forms the basis of the $\Lambda$ Cold Dark Matter ($\Lambda$CDM) model, considered the standard model of cosmology \citep{Weinberg/1989}. However, despite its simplicity, this model presents a significant theoretical challenge, often referred to as the cosmological coincidence problem \citep{CC1,CC2}. These problems keep motivating the study of extensions of general relativity, whereby many different theories and approaches have been proposed and considered \citep{MG1,MG2}.

Modified gravity is an effective approach to describe the acceleration expansion of the universe, except for the introduction of inflation and/or dark energy components. Among the various approaches to modified gravity, a particularly intriguing class is based on the formulation of nonmetricity. In particular, starting from the simplest nonmetricity gravity, namely the symmetric teleparallel equivalent of general relativity (STEGR), one can construct modifications such as the $f(Q)$ gravity. Modified gravity theories are increasingly prominent in modern cosmology due to their effectiveness in representing diverse cosmological phenomena. A considerable amount of research has been dedicated to these theories, aiming to address some of the most challenging questions in today's cosmological landscape \cite{RL, BBN, FR, ZH, SM,SM2,lav1, lav2, JIM, HAR, LH, Arora,lss,Ghosh1,Ghosh2, SM3, gh1, gh2,gh3,Hu/2022}.  Furthermore, several studies have investigated the existence and polarization of gravitational waves within the framework of symmetric teleparallel theories of gravity and their modifications, including $f(Q)$ and $f(Q,B)$ theories \cite{GW1,GW2,GW3,GW4,GW5}. These works aim to understand how such modifications affect the properties and dynamics of gravitational waves. The study by Gomes et al. \cite{Gomes/2024} highlights the presence of ghosts and strong coupling issues in $f(Q)$ theories, particularly in cosmological settings, raising concerns about their physical viability. The ghost mode can be eliminated by imposing specific constraints on it \cite{gh1,gh2,gh3}. Apart from this, the nonmetricity family of gravitational theory is an alternative to theories like curvature and torsion-based gravity [see some review reports on these theories \citep{fr1,fr2,fr21, fr22, fr3,ft1,ft2}].

Additionally, in order to acquire a dynamical dark energy sector and early universe exponential acceleration (known as inflation) within the framework of nonmetricity gravity, we need to include the quintessence scalar field Lagrangian in the nonmetricity action, given by 
\begin{equation}
\mathcal{L}_{\phi} = -\frac{1}{2} g^{\mu\nu} \partial_{\mu}\phi\, \partial_{\nu}\phi - V(\phi),
\end{equation}
where the first term represents the kinetic energy of the scalar field \(\phi\), and \(V(\phi)\) is its potential energy, as discussed in \citep{Caldwell/1998,Bahamonde/2018}. The corresponding gravitational field equation is also derived, as outlined in Section \ref{section 2}. 

Choosing a quintessence scalar field model requires selecting a suitable potential \( V(\phi) \) to drive the various dynamical phases of the universe. Various scalar field potentials have been extensively studied in the literature, including quadratic free-field potentials \citep{FF1,FF2}, power-law potentials \citep{PP1,PP2}, exponential potentials \cite{exp}, and hyperbolic potentials \citep{Obs1}. Recently, Heisenberg et al. \citep{exp} investigated the observational implications of future surveys on quintessence models with \( V(\phi) \sim e^{-\lambda\phi} \), placing constraints on the parameter \( \lambda \). Yang et al. \citep{Obs1} also explored a variety of general potentials, such as exponential, hyperbolic, and power-law types, using data from baryon acoustic oscillations (BAO), cosmic microwave background observations (CMB), joint light curve analysis (JLA), redshift space distortions (RSD), and the cosmic chronometers (CC) to constrain cosmological models. More recently, the power-law potential has been applied to determine cosmological parameters through HII starburst galaxy magnitude data and other observational measurements \citep{Obs2}. So far in the literature, researchers have considered various scalar field potentials to explore cosmological scenarios in theoretical and observational studies in the context of gravitational theories, as discussed previously. However, it will be interesting to reconstruct the analytical form of scalar field potential without prior assumptions and based on observational measurements instead of assuming any arbitrary forms of scalar field potential. 

In this work, we select a quintessence scalar field model and attempt to reconstruct the scalar field potential $V(\phi)$ directly from the observational Hubble dataset within the context of the power-law nonmetricity model. This approach allows us to obtain a suitable quintessence scalar field model that aligns with the observational Hubble data under the framework of power-law nonmetricity gravity. The reconstruction is performed using the Gaussian process (GP) method, originally developed by Seikel et al. \citep{Seikel/2012}. Recently, Jesus et al. \citep{RP1} used $H(z)$ and Type Ia supernovae data, while Elizalde et al. \cite{RP2} utilized 40 Hubble data points to reconstruct the dark energy potential via the Gaussian process in a model-independent manner. Similarly, Niu et al. \citep{RP3} employed three data sets-cosmic chronometers (CC), baryon acoustic oscillations (BAO), and a combination of CC+BAO-along with two priors from Planck 2018 and the Nine-Year Wilkinson Microwave Anisotropy Probe (WMAP). They demonstrated how different priors and data sets affect the reconstruction results of the dark energy potential using the Gaussian process. Cai et al. \citep{Cai/2020} applied a model-independent Gaussian process technique to reconstruct the functional forms of \( f(T) \) directly from observational data, while Gadbail et al. \citep{Gadbail/2024} and Yang et al. \citep{Yang/2024} employed the same technique to reconstruct the functional forms of \( f(Q) \) from observational data. Apart from these, some other works have been explored for various dark energy scenarios using the same approach; for instance, see \cite{jls1,jls2,jls3}.

The structure of this paper is as follows: In Sections \ref{section 2} and \ref{section 3}, we provide a brief overview of modified symmetric teleparallel gravity in the presence of a scalar field, as well as its implications for FLRW cosmology. In Section \ref{section 4}, we introduce the observational Hubble dataset and employ Gaussian process methods to reconstruct the Hubble function and its first-order derivative. Section \ref{section 5} details the step-by-step reconstruction of the scalar field potential using the OHD dataset within the framework of the power-law nonmetricity model, and we compare our reconstructed model with other scalar field potentials. Finally, in Section \ref{section 6}, we discuss our results and their implications.\\

\section{\NoCaseChange{nonmetricity Gravitational Theory}}
\label{section 2}
In this section, we present a brief overview of modified symmetric teleparallel gravity in the presence of a scalar field. This theory is derived from metric-affine geometry with vanishing curvature and torsion, where the nonmetricity tensor $Q_{\sigma\mu\nu}$ plays a crucial role and it is defined as $Q_{\sigma\mu\nu}=\nabla_{\sigma}g_{\mu\nu}$. Furthermore, the presence of the scalar field introduces new degrees of freedom and allows for a more prosperous dynamic behavior, particularly in cosmological contexts. \\
 The action of nonmetricity gravitational theory in the presence of a scalar field is defined as
\begin{equation}
\label{5}
S=\int \left\{-\frac{1}{2\kappa^2}\left[Q+f(Q)\right]+\mathcal{L}_{\phi}+\mathcal{L}_m\right\}\sqrt{-g}\,d^4x.
\end{equation}
Here $\mathcal{L}_{\phi}=-\frac{1}{2}g^{\mu\nu}\partial_{\mu}\phi\,\partial_{\nu}\phi-V(\phi)$ \citep{Caldwell/1998,Bahamonde/2018}, where $V(\phi)$ is the scalar field potential. $\mathcal{L}_m$ is the matter Lagrangian, $f(Q)$ represents any function of the scalar $Q$, and $g$ denotes the determinant of $g_{\mu\nu}$.\\
The nonmetricity scalar $Q$ is given by
\begin{equation}
\label{4}
Q=-Q_{\sigma\mu\nu}P^{\sigma\mu\nu},
\end{equation} 
where $P_{\,\,\mu\nu}^{\sigma}$ is the superpotential tensor, and it can be written as 
\begin{equation}
\label{3}
4P_{\,\,\mu\nu}^{\sigma}=-Q^{\sigma}_{\,\,\,\,\mu\nu}+2Q^{\,\,\,\,\,\,\sigma}_{(\mu\,\,\,\,\nu)}-Q^{\sigma}g_{\mu\nu}-\tilde{Q}^{\sigma}g_{\mu\nu}-\delta^{\sigma}_{(\mu}\, Q\,_{\nu)},
\end{equation} 
with $ Q_{\sigma}=Q_{\sigma\,\,\,\,\mu}^{\,\,\,\,\mu}$, $\tilde{Q}_{\sigma}=Q^{\mu}_{\,\,\,\,\sigma\mu}$ are the traces of nonmetricity tensor.\\
Varying action \eqref{5} with respect to the metric, the gravitational field equation for $f(Q)$ is obtained as
\begin{multline}
\label{7}
\frac{2}{\sqrt{-g}}\nabla_{\sigma}\left((1+f_Q)\sqrt{-g}\,P^{\sigma}_{\,\,\mu\nu}\right)+\frac{1}{2}(Q+f(Q))\,g_{\mu\nu}
+\\(1+f_Q)\left(P_{\mu\sigma\lambda}Q_{\nu}^{\,\,\,\sigma\lambda}-2Q_{\sigma\lambda\mu}P^{\sigma\lambda}_{\,\,\,\,\,\,\nu}\right)= T_{\mu\nu}^{\phi}+T_{\mu\nu},
\end{multline}
where $f_Q=\frac{d f}{d Q}$. The energy-momentum tensor for scalar field and matter are defined as
\begin{eqnarray}
    T_{\mu\nu}^{\phi}&\equiv & \partial_{\mu}\phi\,\partial_{\nu}\phi-\frac{1}{2}g_{\mu\nu}\,g_{\alpha\beta}\,\partial^{\alpha}\phi\,\partial^{\beta}\phi-g_{\mu\nu}V(\phi),\\
    T_{\mu\nu}&\equiv & -\frac{2}{\sqrt{-g}}\frac{\delta(\sqrt{-g})\mathcal{L}_m} {\delta g^{\mu\nu}},
\end{eqnarray}
respectively. On the other hand, the connection field equation is 
\begin{equation}
\nabla_{\mu}\nabla_{\nu} \left(f_{Q}\sqrt{-g}\,P_{\,\,\,\,\,\,\lambda}^{\mu\nu}\right)=0.
\end{equation}
\section{\NoCaseChange{FLRW cosmology}}
\label{section 3}
In a subsequent analysis, we consider a flat, homogeneous, and isotropic universe, described by the Friedmann-Lemaitre-Robertson-Walker (FLRW) spacetime metric, which is expressed as:
\begin{equation}
ds^2 = -dt^2 + a^2(t)\, \delta_{ij}\, dx^i\, dx^j, \,\,\,\,\, (i,j=1,2,3),
\end{equation}
where \( a(t) \) represents the cosmological scale factor. The corresponding nonmetricity scalar, \( Q \), is derived as \( Q = 6H^2 \), with \( H = \frac{\dot{a}}{a} \) being the Hubble parameter. Here, the dot indicates differentiation with respect to the time coordinate \( t \). \\
Substituting the FLRW metric into the general field equation \eqref{7}, we obtain the relevant Friedmann equations for \( f(Q) \) cosmology as
 
\begin{equation}
\label{f1}
3H^2=\rho_m+\rho_{f}+\rho_{\phi},
\end{equation} 
\begin{equation}
\label{f2}
2\dot{H}+3H^2=-(p_m+p_{f}+p_{\phi}).
\end{equation}
Here 
\begin{eqnarray}
\label{12}
    \rho_{f}&=&\frac{f}{2}-Q\,f_Q,\\
    p_{f}&=&2\dot{H}(2Q\,f_{QQ}+f_Q)-\rho_{f},
\end{eqnarray}
are the energy density and pressure contributed by the modified part of geometry, and
\begin{eqnarray}
\label{12}
    \rho_{\phi}&=&\frac{1}{2}\dot{\phi}^2+V(\phi),\\
    p_{\phi}&=&\frac{1}{2}\dot{\phi}^2-V(\phi)\label{13}
\end{eqnarray}
are the energy density and pressure corresponding to the scalar field.\\
Additionally, the conservation equation of matter fluid, dark energy, and scalar field,
\begin{eqnarray}
\dot{\rho}_{m}+3H(\rho_{m}+p_{m})&=&0,\\
\dot{\rho}_{f}+3H(\rho_{f}+p_{f})&=&0,\\
\dot{\rho}_{\phi}+3H(\rho_{\phi}+p_{\phi})&=&0,
\label{c}
\end{eqnarray}
respectively.
In this analysis, we concentrate on the late-time behavior of the cosmic fluid, allowing us to disregard radiation and focus solely on the contribution from pressureless matter. As a result, we set the matter pressure \( p_m = 0 \), and the matter density is given by \( \rho_m = 3H_0^2 \, \Omega_{0m}(1+z)^3 \), where the subscript zero refers to values measured at the present epoch. Here \( z \) represents the redshift, which is defined as \( z = \frac{1}{a} - 1 \), with \( a \) being the scale factor.\\
Using the equations from \eqref{f1} to \eqref{13}, along with the expression for the matter density, we can determine the kinetic energy \( T \equiv \frac{1}{2}\dot{\phi}^2 \) and the scalar field potential \( V \) as follows:
\begin{equation}
\label{KE}
    T=-\dot{H}(2Qf_{QQ}+f_Q+1)-\frac{3}{2}H_0^2\,\Omega_{0m}(1+z)^3,
\end{equation}
\begin{multline}
\label{Vz}
    V=(3H^2+Qf_Q-\frac{f}{2})-\frac{3}{2}H_0^2\,\Omega_{0m}(1+z)^3\\+\dot{H}(2Qf_{QQ}+f_Q+1),
\end{multline}
respectively. In this study, we need to express the time dependence of the equation in terms of redshift $z$ to investigate the observational study. This can be achieved using the relation
\begin{equation}
\label{tz}
    \frac{d}{dt} = -(1+z)H(z)\,\frac{d}{dz},
\end{equation}
where $H(z)$ is the Hubble parameter as a function of redshift.
\section{\NoCaseChange{Dataset and Gaussian Processes}}
\label{section 4}
\begin{figure}[]
\includegraphics[scale=0.41]{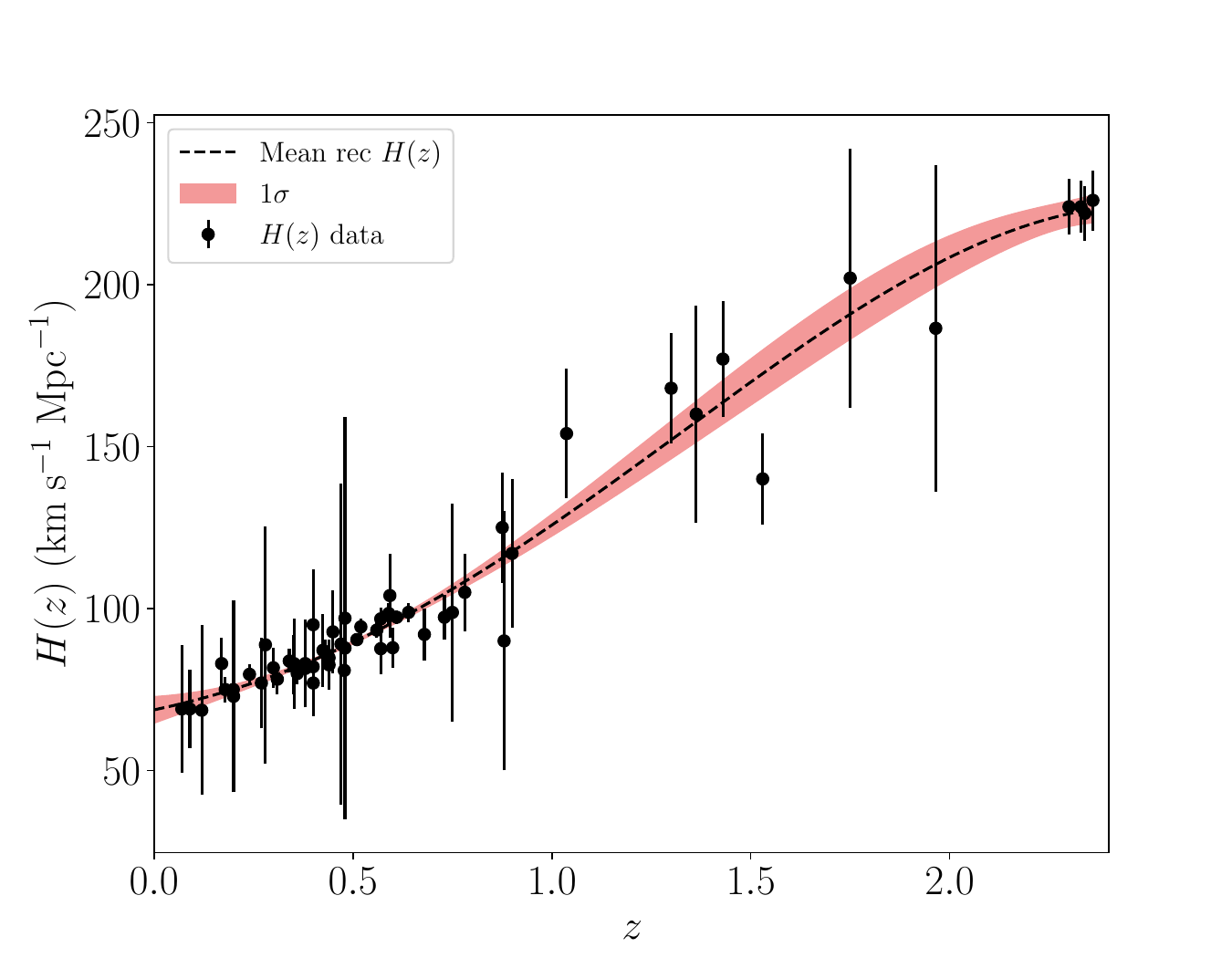}
\includegraphics[scale=0.41]{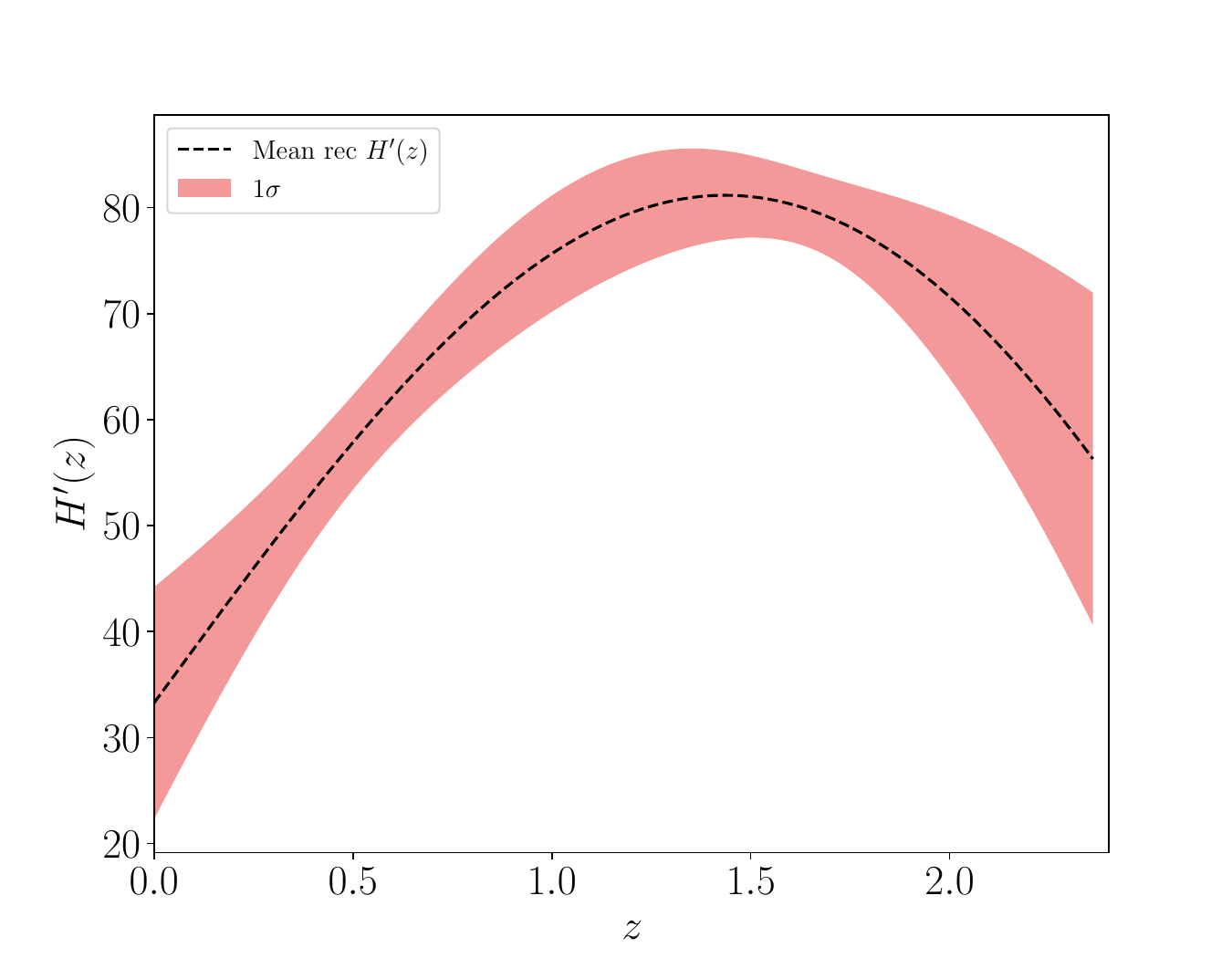}
\caption{\justifying The reconstructed behaviors of $H(z)$ and $H'(z)$ are shown in the upper and lower panels, respectively. These results are obtained from 32 CC data points and 26 BAO data points using the radial method. In both plots, the black dashed line illustrates the mean reconstructed curve, while the shaded area reflects the $1\sigma$ uncertainty attributed to Gaussian process (GP) errors. }
\label{recH}
\end{figure}

This study employs the latest 58 observational data points from the Hubble Space Telescope, incorporating their respective error bars as outlined in Table \ref{Table 1}, to conduct a Gaussian reconstruction. Of these data points, 32 are derived from cosmic chronometer (CC) observations, which independently measure $H(z)$ based on the age progression of passively evolving galaxies. The remaining 26 points are from radial baryon acoustic oscillation (BAO) observations, which gauge galaxy clustering using the BAO peak position as a standard ruler--a measure dependent on the sound horizon and thus model-dependent. Although CC data alone is insufficient to constrain $f(z)$, combining both Hubble samples enhances statistical power, providing more reliable Gaussian process (GP) results. The observational Hubble data (OHD) spans a redshift range of $0.07 < z < 2.42$ across these 58 points.\\
We will apply a nonparametric technique known as Gaussian Processes (GP) \cite{Seikel/2012} to the data set mentioned earlier. This method enables the reconstruction of a continuous function, \( f(x) \), along with its derivatives, from a discrete set of values. Each value in this set is treated as a random variable that follows a Gaussian distribution.\\
The Gaussian process is written as \cite{Seikel/2012,exp1,Mehrabi/2021}
\begin{equation}
    f(x) \mathtt{\sim} \mathcal{GP}\left(\mu(x),k(x,\Tilde{x})\right)
\end{equation}
here, $k(x,\Tilde{x})=\mathbb{E}[(f(x)-\mu(x))(f(\Tilde{x})-\mu(\Tilde{x}))]$ denotes the kernel function, and $x$ represents the observational data points. The term $\mu(x)=\mathbb{E}[f(x)]$ provides the expected mean value of the random variable at each point $ x$. In this study, we utilize the squared exponential function as the kernel to reconstruct both functions and their derivatives, as noted in previous studies \citep{exp1,Mehrabi/2021,exp2}. Known for its flexibility, this kernel serves as a highly adaptable covariance function, expressed as follows

\begin{equation}
    k(x,\Tilde{x})=\sigma^2_f\,exp\left(-\frac{(x-\Tilde{x})^2}{2\,l^2}\right).
\end{equation}
This kernel function is defined by two hyperparameters, $\sigma_f$ and $l$, where $l$ controls the correlation length between neighboring values of $f(x)$, and $\sigma_f$ adjusts the degree of variation in $f(x)$ relative to the mean of the process. \\
In this research, we use the Gaussian Processes in Python (GAPP) package, created by Seikel et al. \citep{Seikel/2012}, to reconstruct the continuous behavior of the Hubble parameter $H(z)$ along with its derivatives, based on observational Hubble data. This model-independent reconstruction is illustrated in Figure \ref{recH}. From our analysis, we obtain a value for the Hubble constant as \( H_0 = 68.71 \pm 4.3 \, \text{km} \, \text{s}^{-1} \, \text{Mpc}^{-1} \), which is consistent with the WMAP observations \cite{WMAP} and the Planck 2015 collaboration value \cite{A8}, within the margin of error. However, we do not need to worry about the $H_0$ tension in this method as it is model-independent and obtained from the dataset directly. \\ 
Further, we used these results to reconstruct the scalar field potential, as discussed in the following section. However, to derive the potential \( V(z) \) from these reconstructions, as shown in equations \eqref{KE} and \eqref{Vz}, a free parameter, \( \Omega_{m0} \), is required. This parameter cannot be directly obtained from the reconstruction and must instead be constrained by other observations, which can provide priors for it. For our analysis, we adopted the prior \( \Omega_{m0} = 0.308\pm 0.012 \), based on the Planck 2015 collaboration \cite{A8}, to ensure reliable results.

\section{\NoCaseChange{Reconstruction of Scalar Field Potential From GPs Utilizing OHD Data}}
\label{section 5}
\begin{figure}[]
\includegraphics[scale=0.41]{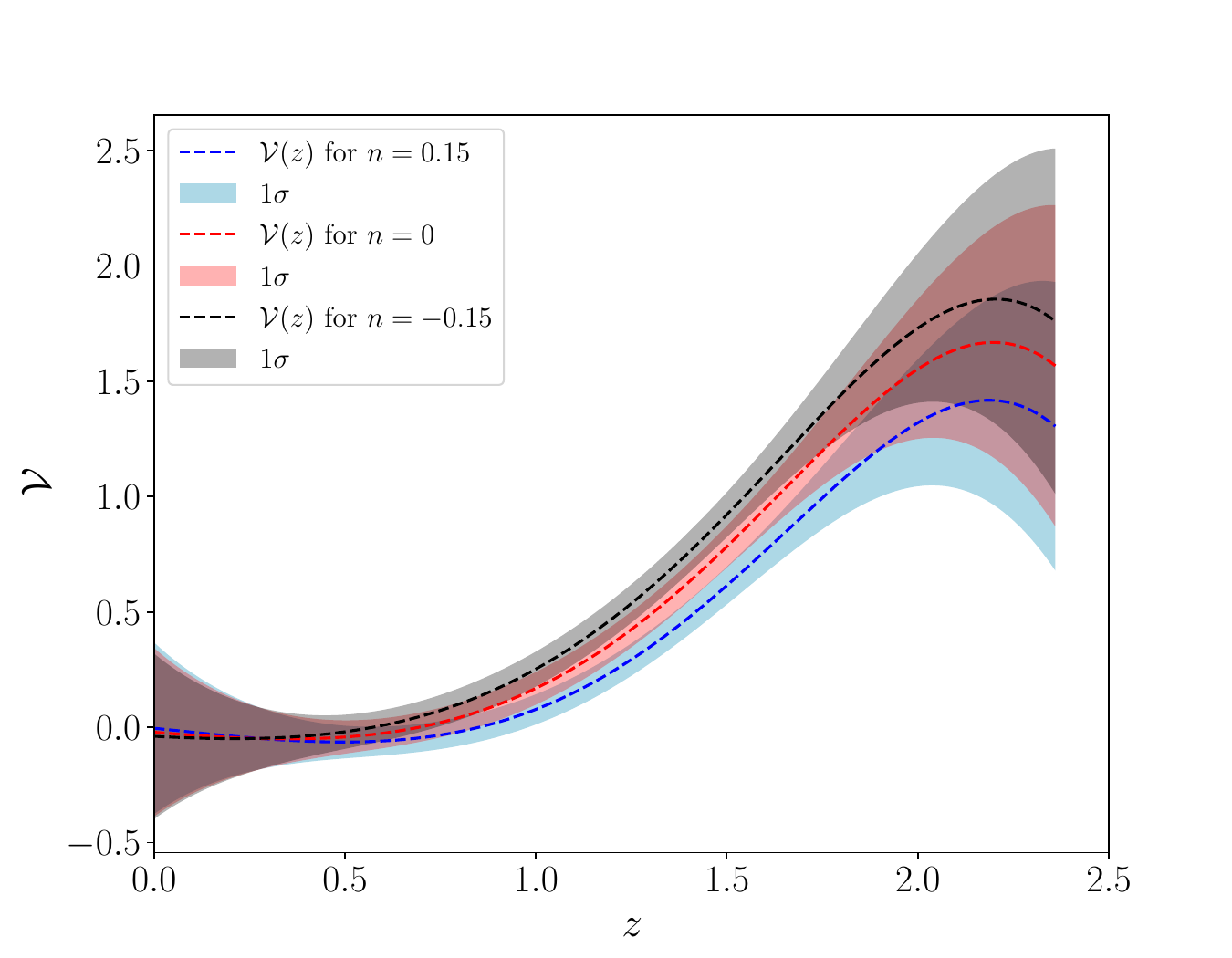}
\includegraphics[scale=0.41]{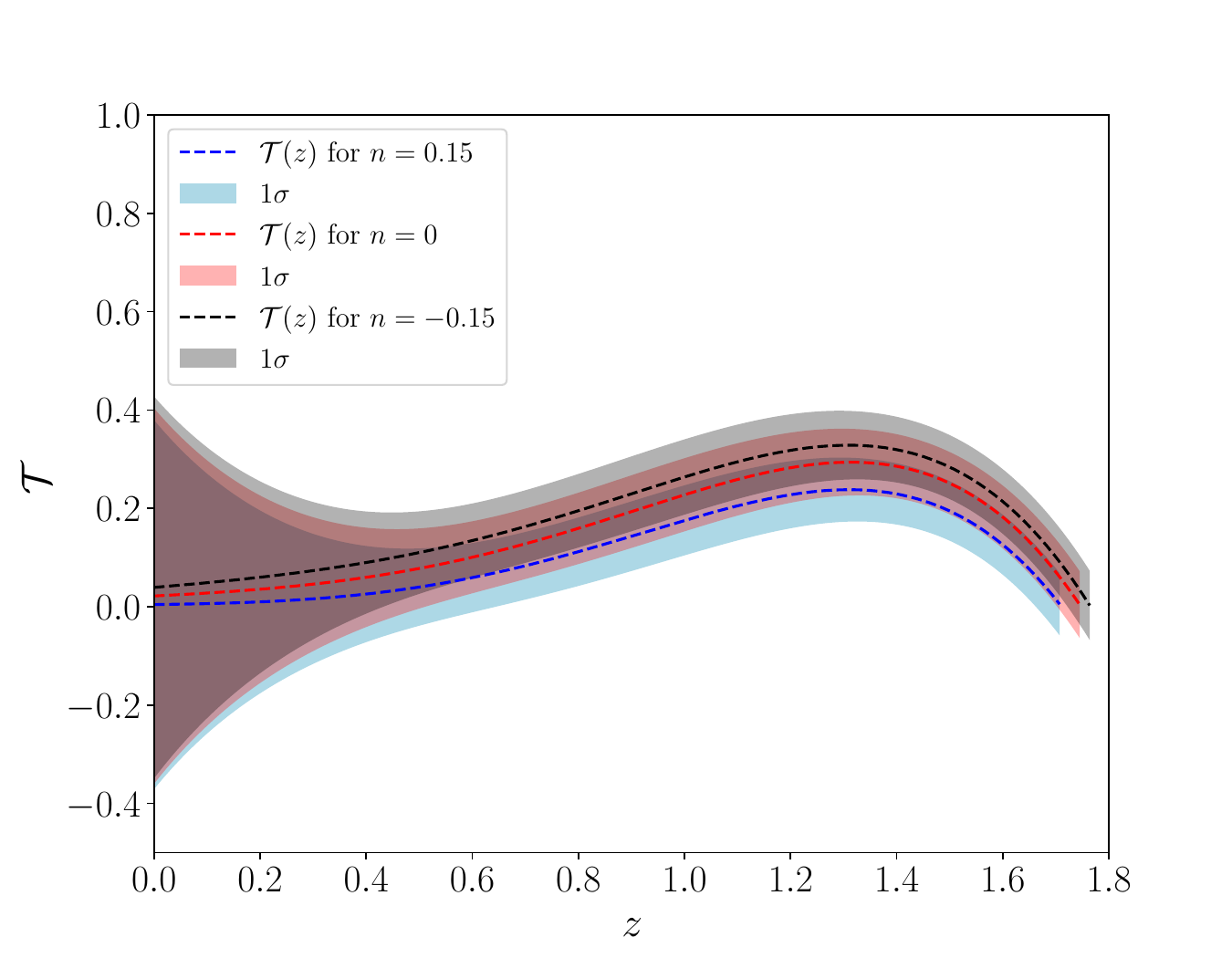}
\caption{\justifying Gaussian process reconstruction of \(\mathcal{V}(z)\) and \(\mathcal{T}(z)\) based on data-driven reconstructions of \(H(z)\) and \(H'(z)\). The dashed lines in both graphs represent the mean reconstructed curves for different values of \(n\), while the lightly shaded regions indicate the \(1\sigma\) uncertainty due to GP errors.}
\label{potentialz}
\end{figure}

\begin{widetext}

\begin{figure}

\begin{minipage}{.33\linewidth}
\centering
\subfloat[]{\label{main:a}\includegraphics[scale=.28]{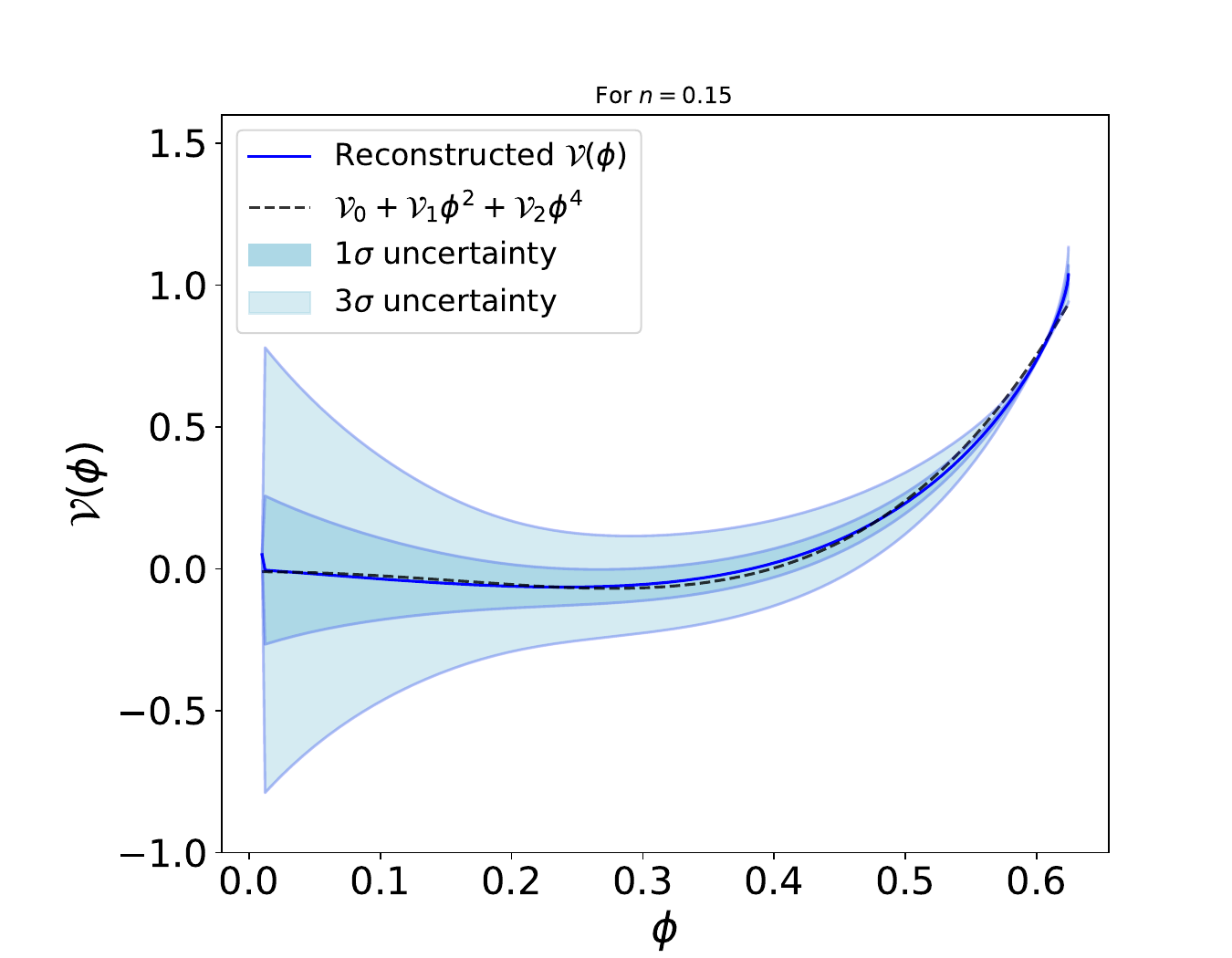}}
\end{minipage}%
\begin{minipage}{.33\linewidth}
\centering
\subfloat[]{\label{main:b}\includegraphics[scale=.28]{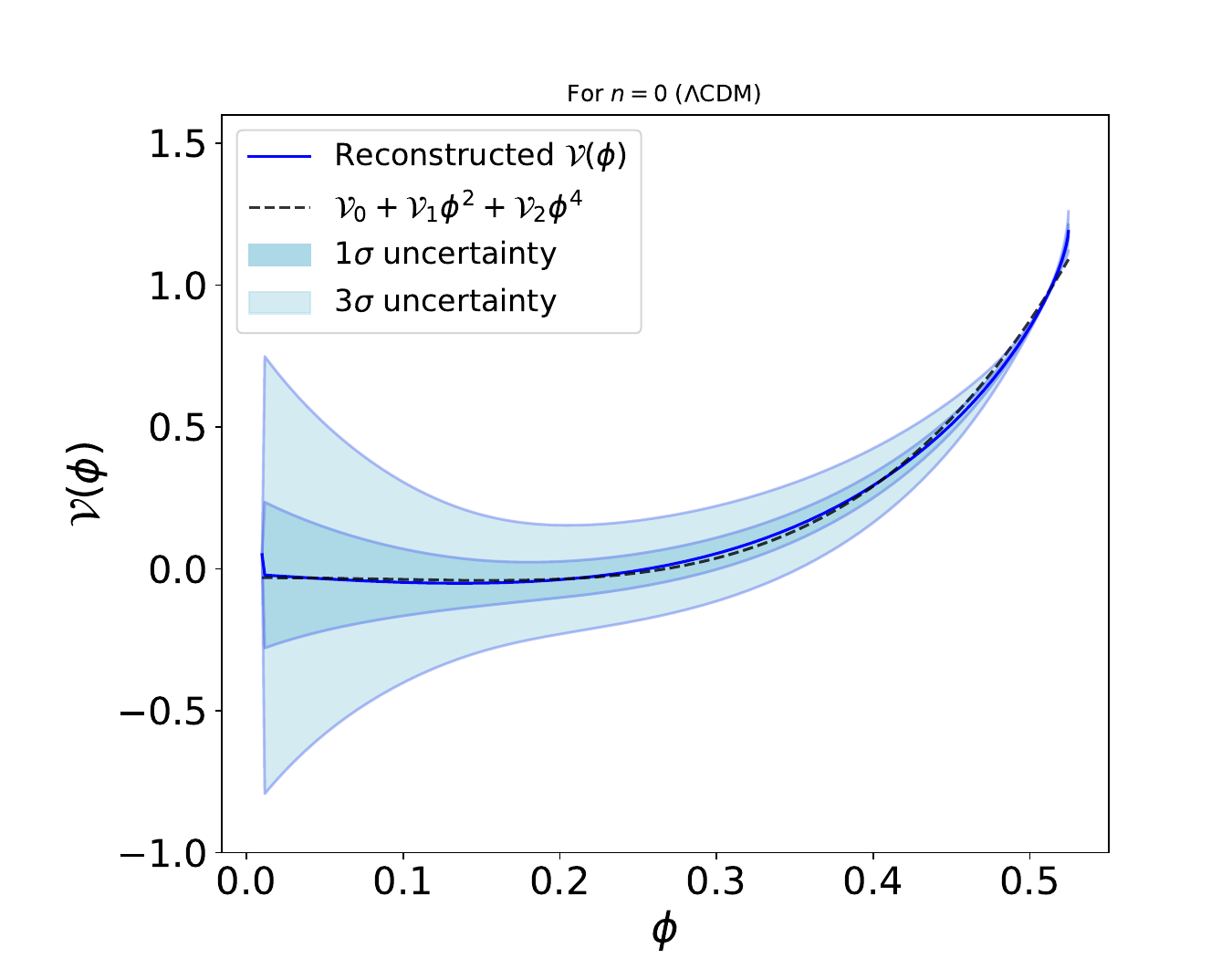}}
\end{minipage}%
\begin{minipage}{.33\linewidth}
\centering
\subfloat[]{\label{main:c}\includegraphics[scale=.28]{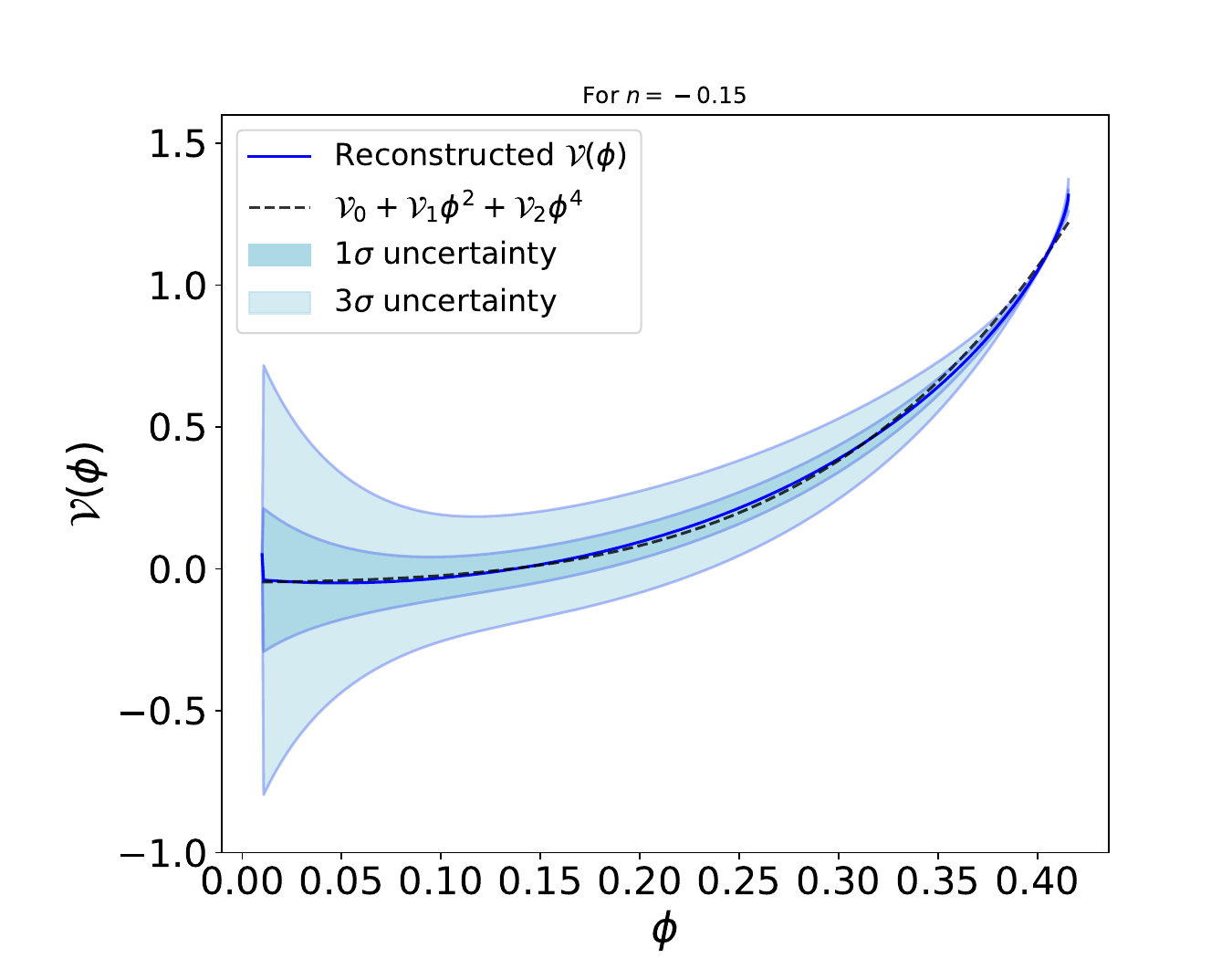}}
\end{minipage}%
\caption{\justifying The reconstructed behavior of the potential \( \mathcal{V}(\phi) \) as a function of the scalar field \( \phi \), derived from the OHD dataset for three different values of the model parameter \( n \), are presented. The blue and light blue regions represent the \( 1\sigma \) and \( 3\sigma \) uncertainties in the reconstruction, respectively, while the black dotted line corresponds to the approximate explicit form of the reconstructed potential.}
\label{potvsphi}
\end{figure}
\end{widetext}

In this section, we reconstruct the scalar field potential corresponding to the power-law nonmetricity model. The functional form of this model is given by $f(Q) = \alpha \left(\frac{Q}{Q_0}\right)^n$, where the parameter $\alpha$ is defined as $\alpha = \frac{(\Omega_{m0}-1)\,6H_0^2}{2n-1}$ and $n$ is a free parameter \cite{RL,JIM,Khyllep/2023}. In the special case where $n = 0$, the model simplifies to $f_{\Lambda \text{CDM}} = -2\Lambda = 6H_0^2(1 - \Omega_{m0})$, which corresponds to the standard $\Lambda$CDM cosmological model, thereby recovering its expansion history for the universe.\\
Using this power law model along with Eq.\eqref{tz}, we find kinetic energy $T$ and scalar field potential $V$ in terms of the Hubble function and its derivative (with respect to $z$) as
\begin{multline}
    T(z)=-\frac{3}{2}H_0^2\,\Omega_{m0}(1+z)^3+(1+z)H'\\\times\left[H+n\,H_0(\Omega_{m0-1})\left(\frac{H}{H_0}\right)^{2n-1}\right],
\end{multline}
and
\begin{multline}
\label{V(z)}
    V(z)=3H^2+3H_0^2(\Omega_{m0}-1)\left(\frac{H}{H_0}\right)^{2n}-\frac{3}{2}H_0^2\,\Omega_{m0}(1+z)^3\\
    -\frac{(1+z)H'}{H}\left[H^2+n\,(\Omega_{m0}-1)H_0^2\left(\frac{H}{H_0}\right)^{2n}\right],
\end{multline}
respectively. Here, the dash represents the derivative with respect to $z$.\\  For numerical analysis, it is beneficial to work with dimensionless quantities. This approach eliminates the need to manage units throughout the calculations, reducing the likelihood of unit mismatches or conversion errors. The dimensionless forms of kinetic energy, $\mathcal{T}(z)$, and potential energy, $\mathcal{V}(z)$, are defined as
\begin{equation}
\label{dimensionlessV}
    \mathcal{T}=\frac{8\pi G}{3H_0^2}T,\,\,\,\,\,\mathcal{V}=\frac{8\pi G}{3H_0^2}V.
\end{equation}
To plot these quantities, we constrained the model parameter $n$ to the range $-0.15 \leq n \leq 0.15$. Extending the parameter beyond $n > 0.15$ causes the kinetic energy of the scalar field to become negative, introducing a ghost field. Ghost fields are characterized by unphysical behavior, such as negative energy densities, which lead to instabilities in the system and violate fundamental physical principles like energy conservation. Therefore, values of $n > 0.15$ are avoided to maintain the physical consistency of the model.\\ 
Similarly, for $n < -0.15$, the potential $\mathcal{V}(\phi)$ takes on imaginary values, which are non-physical and signal the breakdown of the theoretical framework. Imaginary potentials would imply unphysical solutions and introduce mathematical complexities that render the model unsuitable for meaningful cosmological interpretation or further study. Consequently, this lower bound on $n$ ensures that the potential remains real and physically viable.\\
Corresponding to the power-law model, the reconstructed dimensionless scalar field potential and kinetic energy with respect to $z$ are plotted in Fig. \ref{potentialz}.
To clarify, we need to reconstruct the scalar field potential in terms of $\phi$. For this, it is necessary to find the scalar field $\phi$. By applying the chain rule to Eq.\eqref{KE}, we obtain the following expression:
\begin{multline}
\label{Dphi}
    [\phi'(z)]^2=-3\Omega_{m0}(1+z)\left(\frac{H_0}{H}\right)^{2}\\+\frac{2H'}{(1+z)H}\left[1+n\,(\Omega_{m0}-1)\left(\frac{H}{H_0}\right)^{2n-2}\right]
\end{multline}
where primes represent the derivative with respect to redshift $z$. The following step in the application of the GP is to take the approximation of $\phi'$ as
\begin{equation}
\label{20}
    \phi'(z)\approx \frac{\phi(z+\Delta z)-\phi(z)}{\Delta z},
\end{equation}
for small $\Delta z$.
Let's compute the approximate error. We write a Taylor expansion of $\phi(z + \Delta z)$ about $z$, and then we obtain
\begin{equation}
\label{21}
    \phi'(z)=\frac{\phi(z + \Delta z)-\phi(z)}{\Delta z}-\frac{\Delta z}{2}\phi''(\zeta),\,\,\,\,\,\zeta\in (z,z+\Delta z).
\end{equation}
The second term on the right-hand side of Eq.\eqref{21} represents the error term. Since the approximation in Eq.\eqref{20} is derived by omitting this term from the exact expression in Eq.\eqref{21}, this error is referred to as the truncation error. The small parameter \(\Delta z\) indicates the separation between two points, \(z\) and \(z + \Delta z\). As \(\Delta z\) approaches zero, meaning the two points get closer together, the approximation in Eq.\eqref{20} is expected to become more accurate. This holds true if the truncation error vanishes, which happens when \(\phi''(\zeta)\) is well-defined within the interval \((z, z + \Delta z)\). The rate at which the error decreases as \(\Delta z \to 0\) is called the rate of convergence.\\
Using Eq.\eqref{Dphi} along with the approximation for \(\phi'(z)\), we can derive a recursive relation between consecutive redshifts \(z_i\) and \(z_{i+1}\). This allows us to express \(\phi(z_{i+1})\) as a function of \(\phi(z_i)\), along with \(H(z_i)\) and \(H'(z_i)\) as
\begin{multline}
\label{phii}
    \phi(z_{i+1})=\phi(z_i)+(z_{i+1}-z_i)\left[-3\Omega_{m0}(1+z)\left(\frac{H_0}{H}\right)^{2}\right.\\
    \left.+\frac{2H'}{(1+z)H}\left(1+n\,(\Omega_{m0}-1)\left(\frac{H}{H_0}\right)^{2n-2}\right)\right]^{1/2}.
\end{multline}
Using Eqs. \eqref{V(z)}, \eqref{dimensionlessV}, and \eqref{phii}, we numerically plotted the dimensionless scalar field potential \(\mathcal{V}(\phi)\) as a function of scalar field \(\phi\), as shown in Fig. \ref{potvsphi}.

Now, our objective is to determine an explicit functional form for $\mathcal{V}(\phi)$ that closely aligns with the reconstructed $\mathcal{V}(\phi)$ from our analysis.
\\
We adopt a fourth-order polynomial form as
\begin{equation}
    \mathcal{V}(\phi)=\mathcal{V}_0+\mathcal{V}_1\,\phi^2+\mathcal{V}_2\,\phi^4,
\end{equation}
 where $\mathcal{V}_0$, $\mathcal{V}_1$, $\mathcal{V}_2$ are polynomial coefficient. In the reconstruction profile, we presented the mean reconstruction curve alongside the adopted polynomial function. Now, the coefficients are dependent on the model parameter $n$. Here we take the interval for parameter $n$ as
 \begin{equation*}
     n\in[-0.15,0.15]
 \end{equation*}
 and the corresponding polynomial coefficient takes 
 \begin{eqnarray*}
     \mathcal{V}_0&\in &[-0.0463,-0.0069]\\
     \mathcal{V}_1&\in &[-1.5533,1.9601]\\
     \mathcal{V}_2&\in &[10.2105,31.1459].
 \end{eqnarray*}
When we set $n = 0$, our nonmetricity gravitational theory reduces to the standard $\Lambda$CDM cosmological model. This implies that the reconstruction of the potential for $n = 0$ corresponds to the $\Lambda$CDM cosmological model. We have presented the more precise values of the constant coefficients in Table \ref{Table 2}.

Our reconstructed potential is matched with Higgs scalar field potential with $\mathcal{V}_0$ is a constant term, representing the vacuum energy, $\mathcal{V}_1$ is typically associated with the mass term of the scalar field (i.e., $\mathcal{V}_1=\mu^2/2$, where $\mu=m/H_0$) and $\mathcal{V}_2$ is related to the self-interaction of the Higgs field (i.e., $\mathcal{V}_2=\lambda/4$). This potential is commonly used in the Standard Model of particle physics to describe spontaneous symmetry breaking.\\
Our reconstructed potential can be divided into various distinct forms, as discussed below.\\
\textit{Massive free scalar field potential:} The second form is the massive free scalar field potential \(\mathcal{V}(\phi) = \mathcal{V}_1 \phi^2\), where the coefficient \(\mathcal{V}_1\) is defined as \(\mathcal{V}_1 = \mu^2/2\), with \(\mu = m/H_0\) and \(m\) representing the scalar field's mass. This potential is valid only when \(n\) lies between \([-0.08, -0.15]\), where the mass \(m\) falls within the range \(0.5348 H_0\) eV to \(1.9799 H_0\) eV. Outside this range, the free field potential does not align with the reconstructed potential.\\
\textit{Massless self-interacting scalar field potential:} The third form is the massless self-interacting scalar field potential \(\mathcal{V}(\phi) = \mathcal{V}_2 \phi^4\), where \(\mathcal{V}_2 = \lambda/4\) and \(\lambda > 0\) represents the self-interaction coupling constant. This ensures the potential remains bounded from below, maintaining stability. For \(n \in [0.15, -0.15]\), \(\mathcal{V}_2\) is positive, keeping the potential positive-definite and compatible with the \(3\sigma\) region of our reconstructed potential.\\
 \textit{$\phi^2$ potential with vacuum energy:}  Lastly, we have the \(\phi^2\) potential with vacuum energy, \(\mathcal{V}(\phi) = \mathcal{V}_0 + \mathcal{V}_1 \phi^2\), known as Linde's hybrid inflation model. Here, \(\mathcal{V}_0\) represents the constant vacuum energy, while \(\mathcal{V}_1\) corresponds to the mass term, as in the massive free scalar field potential. This form is compatible with the reconstructed potential for \(n \in [-0.08, -0.15]\), similar to the massive free field case.\\
 \textit{Massive self-interacting scalar field potential:} The massive self-interacting scalar field potential is given by 
$\mathcal{V}(\phi) = \mathcal{V}_1 \phi^2 + \mathcal{V}_2 \phi^4$,
where \(\mathcal{V}_1 = -\frac{\mu^2}{2} < 0\) and \(\mathcal{V}_2 = \frac{\lambda}{4} > 0\). Based on these restrictions on \(\mathcal{V}_1\) and \(\mathcal{V}_2\), this potential is consistent with our reconstructed potential only for \(n \in (-0.08, 0.15]\).

\begin{widetext}
    
\begin{table}
\begin{center}
  \caption{The table below specifies the polynomial coefficients with $3\sigma$ error for three different values of the model parameter $n$.}
    \label{Table 2}
    \begin{tabular}{l c c c }
\hline\hline 
Model parameter $n$      & $\mathcal{V}_0$ & $\mathcal{V}_1$ & $\mathcal{V}_2$    \\ \hline\hline 
    $n=0.15$       &  $-0.0069\pm 0.0071$ & $-1.5533\pm 0.0984$ & $10.2105\pm 0.2545$\\[1ex]
    $n=0$ ($\Lambda$CDM)       &  $-0.0307\pm 0.0064$ & $-0.8494\pm 0.1300$ & $17.8900\pm 0.4820$\\[1ex]
$n=-0.15$       &  $-0.0463\pm 0.0055$ & $1.9601\pm 0.1951$ & $31.1459\pm 1.1872$\\[1ex]
\hline
\end{tabular}
\end{center}
\end{table}
\end{widetext}

\section{\NoCaseChange{Conclusions}}
\label{section 6}
In this manuscript, we have focused on reconstructing the early dark energy scalar potential, \( V(\phi) \), using the Gaussian process method within the framework of modified gravity theories, specifically nonmetricity gravity. While previous studies, such as those by Niu et al. \citep{RP1}, Jesus et al. \citep{RP2}, and Elizabeth et al. \citep{RP3}, have explored scalar field dark energy reconstructions, our approach differs in methodology. We have built upon our previous work, Gadbail et al. \citep{Gadbail/2024}, where the Lagrangian function of $f(Q)$ gravity was reconstructed, and Niu's approach, which reconstructed the scalar potential within general relativity. In our study, we have introduced both early-time and late-time dark energy within a unified framework, filling a gap in the literature that typically focuses on either early or late-time dark energy. To perform the reconstruction, we utilized 32 Hubble parameter measurements and 26 BAO measurements. For priors, we applied constraints from Planck 2018 and used the square exponential kernel as the covariance function in the Gaussian process.

To carry out the reconstruction process, we have first reconstructed the Hubble function and its derivatives using the Gaussian process. The presence of both early and late-time dark energy components in our model has enabled us to proceed with reconstructing the scalar field potential. For this, we have considered the widely studied power-law form of the Lagrangian function $f(Q)$ in nonmetricity gravity. As noted in previous studies, the free parameter $n$ has played a crucial role in describing both the present and late-time acceleration of the universe. With this in mind, we have reconstructed the scalar field potential and kinetic energy for different values of $n$. We have then reconstructed the potential \(V(\phi)\) as a function of the scalar field \(\phi\) to compare with the potentials explored in the literature. For better comparison, we have adopted an analytical form for the scalar field potential using a fourth-order polynomial function. Furthermore, we have discussed the parameter space of the coefficients in \(V(\phi)\) to determine the best fit for the reconstructed potential. Additionally, we have compared our results with various other potentials and their parameter values, aiming for better agreement with observational datasets.

In the early phase of the universe, the influence of the reconstructed scalar field potential is high (see Fig. \ref{potentialz}), and it can properly study the early inflation of the universe. At later times, for redshifts $z < 0.15$, both the kinetic and potential energies of the scalar field converge to near zero. This implies that the influence of the scalar field becomes negligible in the present-time evolution of the universe, particularly within the framework of the nonmetricity gravitational theory. As a result, the present-day accelerated expansion of the universe is no longer driven by the scalar field but is instead attributed solely to the effects of the modified geometrical part of nonmetricity gravity. In this context, the modified geometric part of nonmetricity gravity acts as the primary mechanism responsible for the current cosmic acceleration, and the present value of the equation-of-state $w_{f}$ parameter lies in the range $-1.0484\lesssim w_{f}\lesssim-0.9516$ for $n\in[-0.15,0.15]$. In this scenario, the modified part of geometry acts as the source of dark energy. 

This behavior suggests that in the early universe, the dynamics of inflation and matter-dominated expansion were influenced by the scalar field, while in the current epoch, the late-time acceleration is purely governed by the geometrical modifications introduced by the nonmetricity gravity. In future work, we aim to conduct a more rigorous study of the inflationary scenario of the universe by exploring the reconstructed scalar field potential in detail. This analysis will allow us to better understand the role of this potential in driving inflation, providing deeper insights into the early universe's dynamics and the conditions that led to its rapid expansion. Further, it would be interesting to see the same analysis with more data points for the other observational samples, such as SNIa supernova and gamma-ray bursts, which may provide better results.\\

\section*{\NoCaseChange{Data Availability Statement}}
There are no new data associated with this article.

\section*{\NoCaseChange{Acknowledgments}}
GNG acknowledges University Grants Commission (UGC), New Delhi, India for awarding a Senior Research Fellowship (UGC-Ref. No.: 201610122060). SM acknowledges the Japan Society for the Promotion of Science (JSPS) for providing a postdoctoral fellowship. This work of SM and KB is supported by the JSPS KAKENHI Grant (Number: 24KF0100). PKS acknowledges Anusandhan National Research Foundation (ANRF), Department of Science and Technology, Government of India for financial support to carry out Research project No.: CRG/2022/001847 and IUCAA, Pune, India for providing support through the visiting Associateship program. The work of KB is supported by the JSPS KAKENHI Grant Numbers JP21K03547.



\begin{widetext}
\section*{Appendix}

\begin{table*}[hb]
\begin{center}

        \caption{\justifying \textit{Here, table contains the $58$ points of Hubble parameter values $H(z)$ with errors $\sigma _{H}$ from differential age ($32$ points), and BAO and other ($26$ points) approaches, along with references.}}
        \label{Table 1}     
\begin{tabular}{|c c c c c c c c|}\hline
\multicolumn{8}{|c|}{Table-1: $H(z)$ datasets consisting of 57 data points} \\ \hline
\multicolumn{8}{|c|}{CC data (32 points)}  \\ \hline
$z$ & $H(z)$ & $\sigma _{H}$ & Ref. & $z$ & $H(z)$ & $\sigma _{H}$ & Ref. \\ \hline
$0.070$ & $69$ & $19.6$ & \cite{h1} & $0.4783$ & $80$ & $99$ & \cite{h5} \\ \hline
$0.09$ & $69$ & $12$ & \cite{h2} & $0.480$ & $97$ & $62$ & \cite{h1} \\ \hline
$0.120$ & $68.6$ & $26.2$ & \cite{h1} & $0.593$ & $104$ & $13$ & \cite{h3} \\ \hline
$0.170$ & $83$ & $8$ & \cite{h2} & $0.6797$ & $92$ & $8$ & \cite{h3} \\ \hline
$0.1791$ & $75$ & $4$ & \cite{h3} & $0.75$ & $98.8$ & $33.6$ & \cite{h} \\ \hline
$0.1993$ & $75$ & $5$ & \cite{h3} & $0.7812$ & $105$ & $12$ & \cite{h3} \\ \hline
$0.200$ & $72.9$ & $29.6$ & \cite{h4} & $0.8754$ & $125$ & $17$ & \cite{h3} \\ \hline
$0.270$ & $77$ & $14$ & \cite{h2} & $0.880$ & $90$ & $40$ & \cite{h1} \\ \hline
$0.280$ & $88.8$ & $36.6$ & \cite{h4} & $0.900$ & $117$ & $23$ & \cite{h2} \\ \hline  
$0.3519$ & $83$ & $14$ & \cite{h3} & $1.037$ & $154$ & $20$ & \cite{h3} \\ \hline 
$0.3802$ & $83$ & $13.5$ & \cite{h5} & $1.300$ & $168$ & $17$ & \cite{h2} \\ \hline 
$0.400$ & $95$ & $17$ & \cite{h2} & $1.363$ & $160$ & $33.6$ & \cite{h7} \\ \hline 
$0.4004$ & $77$ & $10.2$ & \cite{h5} & $1.430$ & $177$ & $18$ & \cite{h2} \\ \hline 
$0.4247$ & $87.1$ & $11.2$ & \cite{h5} & $1.530$ & $140$ & $14$ & \cite{h2} \\ \hline
$0.4497$ & $92.8$ & $12.9$ & \cite{h5} & $1.750$ & $202$ & $40$ & \cite{h2} \\ \hline
$0.470$ & $89$ & $34$ & \cite{h6} & $1.965$ & $186.5$ & $50.4$ & \cite{h7}  \\ \hline
\multicolumn{8}{|c|}{From BAO \& other method (26 points)} \\ \hline
$z$ & $H(z)$ & $\sigma _{H}$ & Ref. & $z$ & $H(z)$ & $\sigma _{H}$ & Ref. \\ \hline
$0.24$ & $79.69$ & $2.99$ & \cite{h8} & $0.52$ & $94.35$ & $2.64$ & \cite{h10} \\ \hline
$0.30$& $81.7$ & $6.22$ & \cite{h9} & $0.56$ & $93.34$ & $2.3$ & \cite{h10} \\ \hline
$0.31$ & $78.18$ & $4.74$ & \cite{h10} & $0.57$ & $87.6$ & $7.8$ & \cite{h14} \\ \hline
$0.34$ & $83.8$ & $3.66$ & \cite{h8} & $0.57$ & $96.8$ & $3.4$ & \cite{h15} \\ \hline
$0.35$ & $82.7$ & $9.1$ & \cite{h11} & $0.59$ & $98.48$ & $3.18$ & \cite{h10} \\ \hline
$0.36$ & $79.94$ & $3.38$ & \cite{h10} & $0.60$ & $87.9$ & $6.1$ & \cite{h13} \\ \hline
$0.38$ & $81.5$ & $1.9$ & \cite{h12} & $0.61$ & $97.3$ & $2.1$ & \cite{h12} \\ \hline
$ 0.40$ & $82.04$ & $2.03$ & \cite{h10} & $0.64$ & $98.82$ & $2.98$ & \cite{h10}  \\ \hline
$0.43$ & $86.45$ & $3.97$ & \cite{h8} & $0.73$ & $97.3$ & $7.0$ & \cite{h13} \\ \hline
$0.44$ & $82.6$ & $7.8$ & \cite{h13} & $2.30$ & $224$ & $8.6$ & \cite{h16} \\ \hline
$0.44$ & $84.81$ & $1.83$ & \cite{h10} & $2.33$ & $224$ & $8$ & \cite{h17} \\ \hline
$0.48$ & $87.79$ & $2.03$ & \cite{h10} & $2.34$ & $222$ & $8.5$ & \cite{h18} \\ \hline
$0.51$ & $90.4$ & $1.9$ & \cite{h12} & $2.36$ & $226$ & $9.3$ & \cite{h19} \\ \hline
\end{tabular}
 \end{center}
 \end{table*}
\end{widetext}


\begin{thebibliography}{0}%
\makeatletter
\providecommand \@ifxundefined [1]{%
 \@ifx{#1\undefined}
}%
\providecommand \@ifnum [1]{%
 \ifnum #1\expandafter \@firstoftwo
 \else \expandafter \@secondoftwo
 \fi
}%
\providecommand \@ifx [1]{%
 \ifx #1\expandafter \@firstoftwo
 \else \expandafter \@secondoftwo
 \fi
}%
\providecommand \natexlab [1]{#1}%
\providecommand \enquote  [1]{``#1''}%
\providecommand \bibnamefont  [1]{#1}%
\providecommand \bibfnamefont [1]{#1}%
\providecommand \citenamefont [1]{#1}%
\providecommand \href@noop [0]{\@secondoftwo}%
\providecommand \href [0]{\begingroup \@sanitize@url \@href}%
\providecommand \@href[1]{\@@startlink{#1}\@@href}%
\providecommand \@@href[1]{\endgroup#1\@@endlink}%
\providecommand \@sanitize@url [0]{\catcode `\\12\catcode `\$12\catcode `\&12\catcode `\#12\catcode `\^12\catcode `\_12\catcode `\%12\relax}%
\providecommand \@@startlink[1]{}%
\providecommand \@@endlink[0]{}%
\providecommand \url  [0]{\begingroup\@sanitize@url \@url }%
\providecommand \@url [1]{\endgroup\@href {#1}{\urlprefix }}%
\providecommand \urlprefix  [0]{URL }%
\providecommand \Eprint [0]{\href }%
\providecommand \doibase [0]{https://doi.org/}%
\providecommand \selectlanguage [0]{\@gobble}%
\providecommand \bibinfo  [0]{\@secondoftwo}%
\providecommand \bibfield  [0]{\@secondoftwo}%
\providecommand \translation [1]{[#1]}%
\providecommand \BibitemOpen [0]{}%
\providecommand \bibitemStop [0]{}%
\providecommand \bibitemNoStop [0]{.\EOS\space}%
\providecommand \EOS [0]{\spacefactor3000\relax}%
\providecommand \BibitemShut  [1]{\csname bibitem#1\endcsname}%
\let\auto@bib@innerbib\@empty
\end{thebibliography}%


\begin{thebibliography}{90}
\bibitem{A1}  A. G. Riess et al., \href{https://doi.org/10.1086/300499}{ Astron. J. \textbf{116}, 1009 (1998)}.
\bibitem{A2} S. Perlmutter et al., \href{https://doi.org/10.1086/307221}{Astrophys. J. \textbf{517}, 565 (1999)}.

\bibitem{A3} Y. Sofue and V. Rubin, \href{https://doi.org/10.1146/annurev.astro.39.1.137}{ Annu. Rev. Astron. Astrophys. \textbf{39}, 137 (2001)}.
\bibitem{A4} M. Bartelmann and P. Schneider, \href{https://doi.org/10.1016/S0370-1573%2800%2900082-X}{Phys. Rep. \textbf{340}, 291 (2001)}.
\bibitem{A5} D. Clowe, A. Gonzalez, and M. Markevitch, \href{https://doi.org/10.1086/381970}{Astrophys.J. \textbf{604}, 596 (2004)}.
\bibitem{A6} G. Hinshaw et al. (WMAP Collaboration), \href{https://doi.org/10.1088/0067-0049/208/2/19}{Astrophys. J. Suppl. Ser. \textbf{208}, 19 (2013)}.
\bibitem{A7} P. A. R. Ade et al. (Planck Collaboration), \href{https://doi.org/10.1051/0004-6361/201321591}{Astron.
Astrophys. \textbf{571}, A16 (2014)}.
\bibitem{A8} P. A. R. Ade et al. (Planck Collaboration), \href{https://doi.org/10.1051/0004-6361/201525830}{Astron.
Astrophys. \textbf{594}, A13 (2016)}.
\bibitem{A9} N. Aghanim et al. (Planck Collaboration), \href{https://doi.org/10.1051/0004-6361/201833910}{Astron.
Astrophys. \textbf{641}, A6 (2016)}.
\bibitem{A10} A. G. Riess et al., \href{https://doi.org/10.3847/1538-4357/aac82e}{Astrophys. J. \textbf{861}, 126 (2018)}.
\bibitem{DE1} J. Yoo and Y. Watanabe, \href{https://doi.org/10.1142/S0218271812300029}{Int. J. Mod. Phys. D \textbf{21}, 1230002 (2012)}.
\bibitem{DE2} A. A. Costa, R. C. G. Landim, B. Wang, and E. Abdalla,
\href{https://doi.org/10.1140/epjc/s10052-018-6237-7}{Eur. Phys. J. C \textbf{78}, 746 (2018)}.
\bibitem{Weinberg/1989} S. Weinberg, \href{https://doi.org/10.1103/RevModPhys.61.1}{Rev. Mod. Phys. \textbf{61}, 1 (1989)}.
\bibitem{CC1} H. E. S. Velten, R. F. vom Marttens, and W. Zimdahl, \href{https://doi.org/10.1140/epjc/s10052-014-3160-4}{Eur.
Phys. J. C \textbf{74}, 3160 (2014)}.
\bibitem{CC2} N. Sivanandam, \href{https://doi.org/10.1103/PhysRevD.87.083514}{Phys. Rev. D \textbf{87}, 083514 (2013)}.
\bibitem{MG1} T. Clifton, P. G. Ferreira, A. Padilla, and C. Skordis,
\href{https://doi.org/10.1016/j.physrep.2012.01.001}{Phys. Rep. \textbf{513}, 1 (2012)}.
\bibitem{MG2} S. Nojiri, S. D. Odintsov, and V. K. Oikonomou, \href{https://doi.org/10.1016/j.physrep.2017.06.001}{Phys. Rep. \textbf{692}, 1 (2017)}.

\bibitem{RL} R. Lazkoz et al. \href{https://doi.org/10.1103/PhysRevD.100.104027}{Phys. Rev. D \textbf{100}, 104027, (2019)}.

\bibitem{BBN} F. K. Anagnostopoulos et al., \href{https://doi.org/10.1016/j.physletb.2021.136634}{Phys. Lett. B. \textbf{822}, 136634, (2021)}.


\bibitem{FR} N. Frusciante, \href{https://doi.org/10.1103/PhysRevD.103.044021}{Phys. Rev. D \textbf{103}, 044021, (2021)}.

\bibitem{ZH} R. H. Lin, and X. H. Zhai, \href{https://doi.org/10.1103/PhysRevD.103.124001}{Phys. Rev. D \textbf{103}, 124001, (2021)}.

\bibitem{SM}  S. Mandal et al., \href{https://doi.org/10.1103/PhysRevD.102.024057}{Phys. Rev. D \textbf{102}, 024057 (2020)}.

\bibitem{SM2} S. Mandal et al., \href{https://doi.org/10.1103/PhysRevD.102.124029}{Phys. Rev. D \textbf{102}, 124029 (2020)}.

\bibitem{lav1} J. B. Jimenez, L. Heisenberg, and T Koivisto, \href{https://doi.org/10.1103/PhysRevD.98.044048}{Phys. Rev. D \textbf{98}, 044048 (2018)}.

\bibitem{lav2} J. B. Jimenez, L. Heisenberg, and T Koivisto, \href{https://iopscience.iop.org/article/10.1088/1475-7516/2024/03/063}{J. Cosmol. Astropart. Phys. \textbf{03}, 063 (2024)}.


\bibitem{JIM} J. B. Jimenez et al., \href{https://doi.org/10.1103/PhysRevD.101.103507}{Phys. Rev. D \textbf{101}, 103507, (2020)}.

\bibitem{HAR} T. Harko et al., \href{https://doi.org/10.1103/PhysRevD.98.084043}{Phys. Rev. D \textbf{98}, 084043, (2018)}.


\bibitem{LH} L. Heisenberg, \href{https://doi.org/10.1016/j.physrep.2024.02.001}{Phys. Rep. \textbf{1066}, 1-78 (2024)}.
\bibitem{Arora} S. Arora and P.K. Sahoo, \href{ https://doi.org/10.1002/andp.202200233}{Ann. Phys. \textbf{534}, 2200233 (2022)}.
\bibitem{lss} O. Sokoliuk et al., \href{https://doi.org/10.1093/mnras/stad968}{Mon. Not.Roy. Astr. Soc. \textbf{522}, 252-267 (2023)}.
\bibitem{Ghosh1} S. Ghosh. R. Solanki and P.K. Sahoo, \href{https://doi.org/10.1088/1402-4896/ad39b5}{Chinese Phys. C 
 \textbf{48}, 095102 (2024)}.
\bibitem{Ghosh2} S. Ghosh. R. Solanki and P.K. Sahoo, \href{https://doi.org/10.1088/1674-1137/ad50aa}{Phys. Scr. \textbf{99}, 055021 (2024)}.

\bibitem{SM3} S. Mandal and K. Bamba, \href{https://iopscience.iop.org/article/10.1088/1475-7516/2024/11/022/meta}{J. Cosmol. Astropart. Phys. \textbf{2024}, 11 (2024)}.

\bibitem{gh1} K. Hu, M. Yamakoshi, T. Katsuragawa, S. Nojiri and T. Qiu, \href{https://journals.aps.org/prd/abstract/10.1103/PhysRevD.108.124030}{ Phys. Rev. D \textbf{108},  124030 (2023)}.

\bibitem{gh2} S. Nojiri, and S. D. Odintsov, \href{https://onlinelibrary.wiley.com/doi/abs/10.1002/prop.202400113}{ Fortschritte der Physik,
\textbf{72}, 240011 (2024)}.

\bibitem{gh3} S. Nojiri, and S. D. Odintsov, 
\href{https://www.sciencedirect.com/science/article/abs/pii/S2212686424001201}{Phys. Dark Univ., \textbf{45}, 101538 (2024)}.

\bibitem{Hu/2022} K. Hu, T. Katsuragawa, and T. Qiu, \href{https://doi.org/10.1103/PhysRevD.106.044025}{Phys. Rev. D \textbf{106}, 044025 (2022)}.
\bibitem{GW1} M. Hohmann et al.,\href{https://doi.org/10.1103/PhysRevD.99.024009}{Phys. Rev. D \textbf{99},  024009 (2019)}.

\bibitem{GW2} I. Soudi et al., \href{https://doi.org/10.1103/PhysRevD.100.044008}{Phys. Rev. D \textbf{100}, 044008 (2019)}.

\bibitem{GW3} S. Capozzielloa, M. Capriolod, and S. Nojiri, \href{https://doi.org/10.1016/j.physletb.2024.138510}{Phys. Lett. B \textbf{850}, 138510 (2024)}.

\bibitem{GW4} S. Capozzielloa, and M. Capriolod, \href{https://doi.org/10.1016/j.dark.2024.101548}{Phys. Dark Univ. \textbf{45}, 101548 (2024)}.

\bibitem{GW5} S. Capozzielloa, M. Capriolod, and G. Lambiase, \href{https://doi.org/10.1103/PhysRevD.110.104028}{Phys. Rev. D \textbf{110}, 104028 (2024)}. 

\bibitem{Gomes/2024} D.A. Gomes et al., \href{https://doi.org/10.1103/PhysRevLett.132.141401}{Phys. Rev. Lett. \textbf{132}, 141401 (2024)} .






\bibitem{fr1}
T.~P.~Sotiriou and V.~Faraoni,
\href{10.1103/RevModPhys.82.451}{Rev. Mod. Phys. \textbf{82}, 451-497 (2010)}.

\bibitem{fr2}
A.~De Felice and S.~Tsujikawa,
\href{doi:10.12942/lrr-2010-3}{Living Rev. Rel. \textbf{13}, 3 (2010)}.

\bibitem{fr21}
S.~Capozziello and M.~De Laurentis,
\href{doi:10.1016/j.physrep.2011.09.003}{Phys. Rept. \textbf{509}, 167-321 (2011)}.

\bibitem{fr22}
O.~Akarsu, B.~Bulduk, A.~De Felice, N.~Kat\i{}rc\i{} and N.~M.~Uzun,\href{arXiv:2410.23068 [gr-qc]}{arXiv:2410.23068 [gr-qc]}.


\bibitem{fr3}
S.~Nojiri and S.~D.~Odintsov,
\href{10.1016/j.physrep.2011.04.001}{Phys. Rept. \textbf{505}, 59-144 (2011)}.

\bibitem{ft1}
Y.~F.~Cai, S.~Capozziello, M.~De Laurentis and E.~N.~Saridakis,
\href{doi:10.1088/0034-4885/79/10/106901}{Rept. Prog. Phys. \textbf{79}, no.10, 106901 (2016)}.

\bibitem{ft2}
S.~Bahamonde et al.,
\href{doi:10.1088/1361-6633/ac9cef}{Rept. Prog. Phys. \textbf{86}, no.2, 026901 (2023)}.


\bibitem{Caldwell/1998} R.R. Caldwell, R. Dave, and P.J. Steinhardt, \href{https://doi.org/10.1103/PhysRevLett.80.1582}{Phys. Rev. Lett. \textbf{80}, 1582 (1998)}.
\bibitem{Bahamonde/2018} S. Bahamonde et al., \href{https://doi.org/10.1016/j.physrep.2018.09.001}{Phys. Rept. \textbf{775}, 1-122 (2018)}.

\bibitem{FF1} B. Ratra, \href{https://doi.org/10.1103/PhysRevD.44.352}{Phys. Rev. D \textbf{44}, 352 (1991)}.

\bibitem{FF2} L.A. Urena-Lopez and M.J. Reyes-Ibarra, 
\href{https://doi.org/10.1142/S0218271809014674}{Int. J. Mod. Phys. D \textbf{18}, 621 (2009)}.

\bibitem{PP1} P.J.E. Peebles and B. Ratra, \href{https://doi.org/10.1086/185100}{Astrophys.
J. Lett. \textbf{325}, L17 (1988)}.
\bibitem{PP2} B. Ratra and P.J.E. Peebles, \href{https://doi.org/10.1103/PhysRevD.37.3406}{Phys. Rev. D \textbf{37}, 3406 (1988)}.
\bibitem{exp} L. Heisenberg et al., \href{https://doi.org/10.1103/PhysRevD.98.123502}{Phys. Rev. D \textbf{98}, 123502 (2018)}.

\bibitem{Obs1} W. Yang et al., \href{https://doi.org/10.1103/PhysRevD.100.023522}{Phys. Rev. D \textbf{100}, 023522 (2019)}.
\bibitem{Obs2} S. Cao, J. Ryan and B. Ratra, \href{https://doi.org/10.1093/mnras/staa2190}{Mon. Not. Roy. Astron. Soc. \textbf{497}, 3191 (2020)}.
\bibitem{Seikel/2012} M. Seikel, C. Clarkson, and M. Smith, \href{https://doi.org/10.1088/1475-7516/2012/06/036}{J. Cosmol. Astropart. Phys., \textbf{06}, 036, (2012)}.


\bibitem{RP1} J.F. Jesus et al., \href{https://doi.org/10.1088/1475-7516/2022/11/037}{J. Cosmol. Astropart. Phys.,\textbf{ 11}, 037 (2022)}.
\bibitem{RP2} E. Elizalde, M. Khurshudyan, K. Myrzakulov, and S. Bekov, \href{https://doi.org/10.1007/s10511-024-09828-z}{Astrophysics \textbf{67}, 192 (2024)}.
\bibitem{RP3} J. Niu et al., \href{https://doi.org/10.3847/1538-4357/ad5fef}{ Astrophys. J., \textbf{972}, 14 (2024)}
\bibitem{Cai/2020} Y. Cai, M. Khurshudyan1, and E. N. Saridakis, \href{https://doi.org/10.3847/1538-4357/ab5a7f}{ Astrophys. J., \textbf{888}, 62 (2020)}.

\bibitem{Gadbail/2024} G.N. Gadbail, S. Mandal, and P.K. Sahoo, \href{https://doi.org/10.3847/1538-4357/ad5cf4}{Astrophys. J., \textbf{972}, 174 (2024)}.
\bibitem{Yang/2024} Y. Yang et al., \href{https://doi.org/10.1093/mnras/stae1905}{Mon. Not. Roy. Astron. Soc. \textbf{533}, 2232-2241 (2024)}.

\bibitem{jls1} R. C. Bernardo, and J. L. Said  \href{https://doi.org/10.1088/1475-7516/2021/09/014}{J. Cosmol. Astropart. Phys. \textbf{09}, 014 (2021)}.
\bibitem{jls2} R. C. Bernardo, and J. L. Said  \href{https://doi.org/10.1088/1475-7516/2021/08/027}{J. Cosmol. Astropart. Phys. \textbf{08},  027 (2021)}.
\bibitem{jls3}R. C. Bernardo et al.,  \href{https://doi.org/10.1016/j.dark.2022.101017}{Phys. Dark Univ. \textbf{36},  101017 (2022)}.

\bibitem{exp1} C.E. Rasmussen, and C.K. Williams,  \href{https://doi.org/10.7551/mitpress/3206.001.0001}{2005, Gaussian Processes for Machine Learning (Cambridge, MA: MIT Press)}.

\bibitem{Mehrabi/2021} A. Mehrabi, and M. Rezaei, \href{https://doi.org/10.3847/1538-4357/ac2fff}{Astrophys. J. \textbf{923}, 274 (2021)}.
\bibitem{exp2} M. Seikel, and C. Clarkson, \href{https://doi.org/10.48550/arXiv.1311.6678}{2013, arXiv:1311.6678}.

\bibitem{WMAP} C. L. Bennett (WMAP observations) et al., \href{https://doi.org/10.1088/0067-0049/208/2/20}{ ApJS \textbf{208}, 20 (2013)}.

\bibitem{Khyllep/2023} W. Khyllep et al., \href{https://doi.org/10.1103/PhysRevD.107.044022}{Phys. Rev. D \textbf{107}, 044022 (2023)}.









\bibitem{h1} S. D. et al., \href{https://doi.org/10.1088/1475-7516/2010/02/008}
{J. Cosmol. Astropart. Phys. \textbf{02}, 008, (2010)}

\bibitem{h2} S. J.,  L. Verde , and R. Jimenez , \href{https://doi.org/10.1103/PhysRevD.71.123001}%
{Phys. Rev. D \textbf{71}, 123001, (2005)}.

\bibitem{h3} M. Moresco et al., \href{https://doi.org/10.1088/1475-7516/2012/08/006}%
{J. Cosmol. Astropart. Phys. \textbf{08}, 006, (2012)}.
 

\bibitem{h4} Z. Cong et al., \href{https://doi.org/10.1088/1674-4527/14/10/002}%
{Research in Astron. and Astrop. \textbf{14}, 1221, (2014)}.

\bibitem{h5} M. Moresco et al., \href{https://doi.org/10.1088/1475-7516/2016/05/014}{J. Cosmol. Astropart. Phys. \textbf{05}, 014, (2016)}.

\bibitem{h6} A. L. Ratsimbazafy  et al., \href{https://doi.org/10.1093/mnras/stx301}{Mon. Not. Roy. Astron. Soc. \textbf{467}, 3239, (2017)}.
\bibitem{h} N. Borghi, M. Moresco, and A. Cimatti, \href{https://doi.org/10.3847/2041-8213/ac3fb2}{ApJL \textbf{928,} L4 (2022)}.
\bibitem{h7}  M. Moresco, \href{https://doi.org/10.1093/mnrasl/slv037}{Mon. Not. Roy. Astron. Soc. Lett. \textbf{450}, L16, (2015)}.

\bibitem{h8}  E. Gaztaaga et al., \href{https://doi.org/10.1111/j.1365-2966.2009.15405.x}
{Mon. Not. Roy. Astron. Soc. \textbf{399}, 1663, (2009)}.

\bibitem{h9}  A. Oka et al., \href{https://doi.org/10.1093/mnras/stu111}{Mon. Not. Roy. Astron. Soc. \textbf{439}, 2515, (2014)}.

\bibitem{h10}  Y. Wang et al., \href{https://doi.org/10.1093/mnras/stx1090}{Mon. Not. Roy. Astron. Soc. \textbf{469}, 3762, (2017)}.

\bibitem{h11} C. H. Chuang, Wang Y., \href{https://doi.org/10.1093/mnras/stt1290}%
{Mon. Not. Roy. Astron. Soc. \textbf{435}, 255, (2013)}.

\bibitem{h12} S. Alam et al., \href{https://doi.org/10.1093/mnras/stx721}{Mon. Not. Roy. Astron. Soc. \textbf{470}, 2617, (2017)}.

\bibitem{h13} C. Blake  et al., \href{https://doi.org/10.1111/j.1365-2966.2012.21473.x}{Mon. Not. Roy. Astron. Soc. \textbf{425}, 405, (2012)}.

\bibitem{h14} C. H. Chuang et al., \href{https://doi.org/10.1093/mnras/stt988}{Mon. Not. Roy. Astron. Soc. \textbf{433}, 3559, (2013)}.

\bibitem{h15} L.  Anderson et al., \href{https://doi.org/10.1093/mnras/stu523}{Mon. Not. Roy. Astron. Soc. \textbf{441}, 24, (2014)}.

\bibitem{h16}  N. G. Busca et al., \href{https://doi.org/10.1051/0004-6361/201220724}{Astron. Astrophys. \textbf{552}, A96, (2013)}.

\bibitem{h17} J. E. Bautista et al., \href{https://doi.org/10.1051/0004-6361/201730533}{Astron. Astrophys. \textbf{603}, A12, (2017)}.

\bibitem{h18} T. Delubac et al., \href{https://doi.org/10.1051/0004-6361/201423969}{Astron. Astrophys. \textbf{574}, A59, (2015)}.

\bibitem{h19} A. Font-Ribera et al., \href{https://doi.org/10.1088/1475-7516/2014/05/027}{J. Cosmol. Astropart. Phys. \textbf{05}, 027, (2014)}.
\end{thebibliography}
\end{document}